%% file: ACSOS2025.tex
\documentclass[conference]{IEEEtran}
\IEEEoverridecommandlockouts
\usepackage{cite}
\usepackage{amsmath,amssymb,amsfonts}
\usepackage{algorithmic}
\usepackage{graphicx}
\usepackage{textcomp}
\usepackage{xcolor}
\usepackage{subcaption}
\usepackage{hyperref}
\usepackage{tikz}
\usepackage{pgf-umlsd}

\usepackage{pifont}

\usepackage{listings}
\usepackage{xcolor}
\def\BibTeX{{\rm B\kern-.05em{\sc i\kern-.025em b}\kern-.08em
    T\kern-.1667em\lower.7ex\hbox{E}\kern-.125emX}}

\begin{document}

\title{FedLAD: A Modular and Adaptive Testbed for Federated Log Anomaly Detection}

\author{Yihan Liao$^{1}$, Jacky Keung$^{1}$, Zhenyu Mao$^{1,*}$, Jingyu Zhang$^{1}$ and Jialong Li$^{2}$\\
	\normalsize $^{1}$City University of Hong Kong, Hong Kong, China\\
	\normalsize $^{2}$Waseda University, Tokyo, Japan\\
	\normalsize yihanliao4-c@my.cityu.edu.hk, jacky.keung@cityu.edu.hk, jzhang2297-c@my.cityu.edu.hk, lijialong@fuji.waseda.jp\\
	\normalsize *corresponding author: zhenyumao2-c@my.cityu.edu.hk
}

\maketitle

\begin{abstract}
Log-based anomaly detection (LAD) is critical for ensuring the reliability of large-scale distributed systems. However, most existing LAD approaches assume centralized training, which is often impractical due to privacy constraints and the decentralized nature of system logs. While federated learning (FL) offers a promising alternative, there is a lack of dedicated testbeds tailored to the needs of LAD in federated settings. To address this, we present FedLAD, a unified platform for training and evaluating LAD models under FL constraints. FedLAD supports plug-and-play integration of diverse LAD models, benchmark datasets, and aggregation strategies, while offering runtime support for validation logging (self-monitoring), parameter tuning (self-configuration), and adaptive strategy control (self-adaptation). By enabling reproducible and scalable experimentation, FedLAD bridges the gap between FL frameworks and LAD requirements, providing a solid foundation for future research.
Project code is publicly available at: \url{https://github.com/AA-cityu/FedLAD}.
\end{abstract}


\input{1_Introduction}
\input{2_Background} 
\input{3_System_Design}
\input{4_Implementation}
\input{5_Example_Experiment}
\input{6_Threat_to_Validity}
\input{7_Conclusion}

\bibliographystyle{IEEEtran}
\bibliography{ACSOS2025}

\end{document}

%% file: 1_Introduction.tex
\section{Introduction}
\label{sec:introduction}

Modern large-scale software systems, ranging from distributed cloud platforms to edge computing infrastructures, generate massive volumes of system logs that record detailed runtime behaviors~\cite{landauer2023deep}. These logs are critical for diagnosing failures, monitoring performance, and ensuring reliability. Given the scale and complexity, manual inspection is infeasible. As a result, \textbf{L}og-based \textbf{A}nomaly \textbf{D}etection (LAD) has become a vital approach for automatically identifying abnormal behaviors and potential faults~\cite{catillo2022autolog}, and it has been successfully applied in diverse domains such as data centers, Internet of Things, and autonomous vehicles~\cite{le2022log}.

Despite their effectiveness, existing LAD models face several critical limitations. Many assume centralized training with globally accessible logs, which is an impractical assumption due to privacy concerns and organizational data silos. This constraint hinders cross-organizational collaboration, particularly when sensitive operational logs cannot be shared across institutional boundaries~\cite{mothukuri2021federated}. Furthermore, most LAD models are static, struggling to adapt to evolving behaviors or unseen anomalies~\cite{landauer2023deep}. These limitations pose serious obstacles to real-world deployment and reproducible research in distributed settings, where logs are diverse, privacy-sensitive, and continuously evolving~\cite{xie2022loggd}. To overcome the limitations of centralized LAD, recent research has investigated the use of \textbf{F}ederated \textbf{L}earning (FL), which enables multiple clients (e.g., data centers, edge nodes) to collaboratively train LAD models without sharing raw log data, thereby preserving data privacy~\cite{zhang2021survey}. By combining local training with global model aggregation, FL enables knowledge transfer across heterogeneous sources while maintaining data locality, which is an essential property for LAD scenarios involving geographically distributed logs.

While prior works~\cite{hu2025adapting, domini2024proximity} have proposed FL-based LAD methods, they are typically designed for specific experimental setups with tightly integrated implementations, which limit reproducibility, extensibility, and broader reuse. These efforts often focus on particular LAD models or datasets, making it difficult to experiment with alternative architectures or strategies. By contrast, general-purpose FL frameworks such as Flower~\cite{beutel2020flower} provide modular infrastructure for standard domains like image classification. However, they lack native support for structured log data, including log-specific preprocessing or anomaly-centric adaptations. Consequently, there remains a clear gap for a reusable, log-aware FL platform that supports LAD-specific challenges while enabling reproducible and adaptive experimentation. FedLAD fills this gap as the first modular testbed tailored to LAD under FL constraints, integrating pluggable LAD models, log-specific preprocessing, non-independent and identically distributed (IID) simulation, and adaptation driven by detection quality.

Built on a modular architecture, FedLAD decouples key components, including models, datasets, training workflows, and aggregation logic, via standardized interfaces to facilitate extensibility, interoperability, and fair comparison.
In addition, FedLAD incorporates a suite of self-managing features, including self-monitoring for logging per-round performance metrics; self-configuration for recommending hyperparameters; and a self-adaptive executor for triggering strategy switching or early stopping. By lowering the barrier to prototyping, automated evaluation, and training adaptation, FedLAD serves as a unified and reliable testbed for accelerating LAD research in FL.
Detailed enhancements and contributions include:

\begin{itemize}
    \item \textbf{Novel LAD-specific FL testbed}: FedLAD is the first testbed tailored for LAD under FL, bridging the gap between generic FL frameworks and LAD-specific needs.

    \item \textbf{Modular architecture}: All components, including models, datasets, aggregation strategies, and adaptation policies, are decoupled via standardized interfaces, enabling flexible deployment into diverse LAD workflows.
    
    \item \textbf{Experiment extensibility}: FedLAD supports benchmark datasets (BGL~\cite{oliner2007supercomputers}, HDFS~\cite{xu2009detecting}, Thunderbird~\cite{oliner2007supercomputers}), LAD models (DeepLog~\cite{du2017deeplog}, LogAnomaly~\cite{meng2019loganomaly}, NeuralLog~\cite{le2021log}), and aggregation strategies (FedAvg~\cite{mcmahan2017communication}, Scaffold~\cite{karimireddy2020scaffold}, FedAdam~\cite{reddi2020adaptive}, FedProx~\cite{li2020federated}) under both IID and non-IID settings~\cite{zhao2018federated}, while offering unified interfaces for adding new components.
    
    \item \textbf{Trackable training process via self-monitoring}: Per-round validation metrics are logged to enable continuous performance tracking and data-driven runtime decisions.
    
    \item \textbf{Automatic training via self-configuration}: An optional auto-configurator suggests initial hyperparameters based on dataset traits, reducing manual tuning effort.
    
    \item \textbf{Adaptive training via self-adaptation}: A built-in executor monitors runtime feedback and triggers strategy switching or early stopping to improve robustness.
\end{itemize}

%% file: 2_Background.tex
\section{Background}
\label{sec:background}

\subsection{Log Anomaly Detection}

As a foundational technique for preserving software reliability and security, LAD leverages execution logs that encapsulate rich temporal and semantic patterns.
These logs serve as a vital source of information for identifying system faults, diagnosing performance bottlenecks, and detecting security violations.

Recent advances in LAD have embraced deep learning to automatically learn complex sequential and contextual representations from raw log data. 
For example, DeepLog \cite{du2017deeplog} uses LSTMs to model log sequences, LogAnomaly \cite{meng2019loganomaly} combines semantic and sequential features, and LogBERT \cite{guo2021logbert} leverages pretrained language models for contextual understanding.

Despite growing progress in LAD model development, evaluating these methods in a realistic and reproducible manner remains challenging. Real-world log environments vary widely in format, structure, and semantic patterns, making it difficult to compare models fairly or assess their robustness \cite{landauer2023deep}. These challenges underscore the critical need for a configurable testbed that supports diverse datasets, models, and evaluation setups with consistent monitoring and repeatability.

\subsection{Federated Learning-based Log Anomaly Detection}

FL has emerged as a promising paradigm for privacy-preserving machine learning, particularly in domains where data is sensitive and decentralized.
In the context of LAD, FL enables multiple clients, such as data centers, edge nodes, or industrial devices, to collaboratively train anomaly detection models without sharing raw log data.
This is particularly valuable given the sensitive nature of operational and user logs, and the growing regulatory emphasis on data privacy.

Recent studies have demonstrated the feasibility and potential of applying FL to LAD.
For example, FedLog \cite{li2022federated} proposes a communication-efficient FL framework tailored for industrial IoT logs, incorporating a masking mechanism to protect sensitive tokens during training.
FLOGCNN \cite{guo2021anomaly} introduces a lightweight CNN-based model that balances detection performance and model efficiency across distributed clients.
These approaches show that FL can successfully support LAD in decentralized environments while maintaining data privacy.

Despite these advancements, several challenges remain.
Most existing FL-based LAD solutions are tightly coupled to specific model architectures or datasets, limiting their extensibility.
They often prioritize privacy and accuracy metrics, while overlooking key system-level concerns such as runtime adaptivity and client heterogeneity, where clients may differ significantly in log semantics, data volume, or resource availability.
Moreover, common federated aggregation strategies such as FedAvg, Scaffold, and FedAdam remain underexplored, particularly under non-IID data distributions.
Finally, most prior implementations assume static training configurations, reducing the robustness in dynamic environments.

These open challenges call for a unifying testbed that enables flexible experimentation and adapts to dynamic conditions. Therefore, this paper presents FedLAD, a modular, extensible, and self-adaptive testbed that enables scalable and reproducible LAD research within FL.

%% file: 3_System_Design.tex
\section{System Design}
\label{sec:design}

FedLAD is structured to support configurable, component-wise experimentation across LAD models, datasets, and federated strategies. This section introduces its architectural decomposition and key mechanisms for training orchestration, evaluation logging, and adaptation triggering.

\subsection{FedLAD Architecture}

\begin{figure}[t]
    \centering
    \includegraphics[width=1\linewidth]{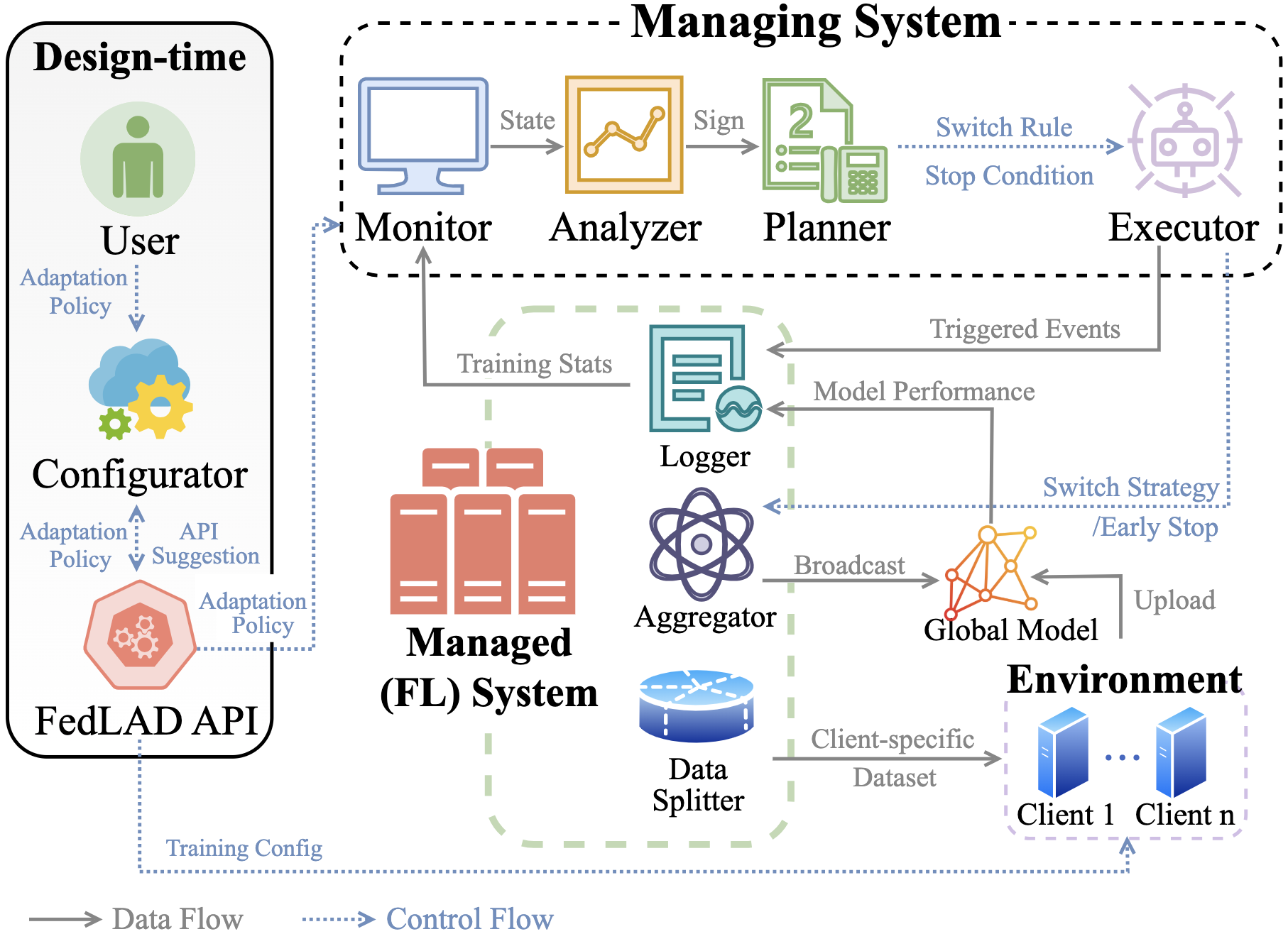}
    \caption{Overview of the architecture of FedLAD.}
    \label{fig:overview}
\end{figure}

As shown in Fig.~\ref{fig:overview}, FedLAD adopts a modular architecture aligned with self-adaptive system principles. Each module encapsulates a distinct role in the federated LAD lifecycle, while the design separates developer-accessible components from runtime execution and coordination. This enables flexible experimentation with LAD models, strategies, and datasets.

FedLAD is organized into four conceptual regions, \textbf{Design-time}, \textbf{Managing System}, \textbf{Managed (FL) System}, and the \textbf{Environment}. Interactions among these layers are governed by well-defined \textit{data flows} (gray) and \textit{control flows} (blue dotted lines), enabling a cohesive testbed for the systematic design and dynamic execution of adaptive FL workflows.

In the design-time region, the user defines both adaptation policies and training configurations via the configurator. The configurator supports manual YAML editing and automated configuration, offering suggestions (e.g., client count) based on dataset characteristics. These settings are then passed to the FedLAD API, which serves as a unified coordination layer between developer-defined inputs and runtime modules. Specifically, adaptation policies are routed to the Managing System, while training configurations are directed to the Environment (i.e., the client-side runtime). As a testbed, FedLAD enables the rapid deployment of adaptation policies and flexible reconfiguration of federated environments, supporting seamless integration of new models, aggregation rules, or adaptation strategies. This design facilitates realistic and controllable experimentation across diverse FL scenarios.

The Managing System implements the core Monitor–Analyzer–Planner–Executor (MAPE) feedback loop that drives self-adaptation. The Monitor collects training states, such as loss and F1-score, from the Managed System. In FedLAD’s implementation, the Analyzer and Planner functionalities are integrated into the Executor module. This unified Executor detects performance trends (e.g., validation plateaus), evaluates adaptation conditions (e.g., early stop or strategy switch), and enforces control decisions back to the training system. By consolidating adaptation logic into a single module, FedLAD simplifies runtime coordination while maintaining explainability and extensibility, enabling rapid prototyping of adaptive strategies without modifying training components.  

The Managed System includes all components necessary for training. The data splitter simulates realistic data heterogeneity across clients, supporting both IID and non-IID partitioning strategies. The aggregator receives model updates from clients and performs aggregation using FedAvg, FedProx, Scaffold, or FedAdam. After aggregation, the Global Model is broadcast back to clients. The Logger captures round-level metrics such as loss and F1-score, and records adaptation events triggered by the Executor. It streams this information to the Monitor as the entry point of the MAPE loop, which processes the data through analysis and planning stages before the Executor triggers adaptive actions.

The Environment represents the client-side runtime, where simulated clients operate on partitioned datasets and perform local training independently. Each client maintains its local model, trains on private data, and uploads updates to the aggregator. Clients also respond to adaptation signals (i.e., early stop and strategy change instructions) broadcast from the Executor. This separation allows the Environment to be flexibly reconfigured to mimic real-world federated scenarios.

In summary, FedLAD’s architecture follows self-adaptive system principles: the Managing System (MAPE loop) observes and steers the Managed System (server-side coordination) and interacts with the Environment (clients) via a defined Coordination Interface (FedLAD API). Beyond runtime adaptation, FedLAD supports design-time configurability through a developer-facing configurator, allowing researchers to flexibly define training parameters, client settings, and adaptation logic. This design enables fast prototyping and controlled experimentation across diverse client environments. All data and control flows are explicitly annotated to ensure traceability and reproducibility. Implementation details are in Section~\ref{sec:implementation}.

\subsection{System Workflow}
\begin{figure}
    \centering
    \includegraphics[width=0.8\linewidth]{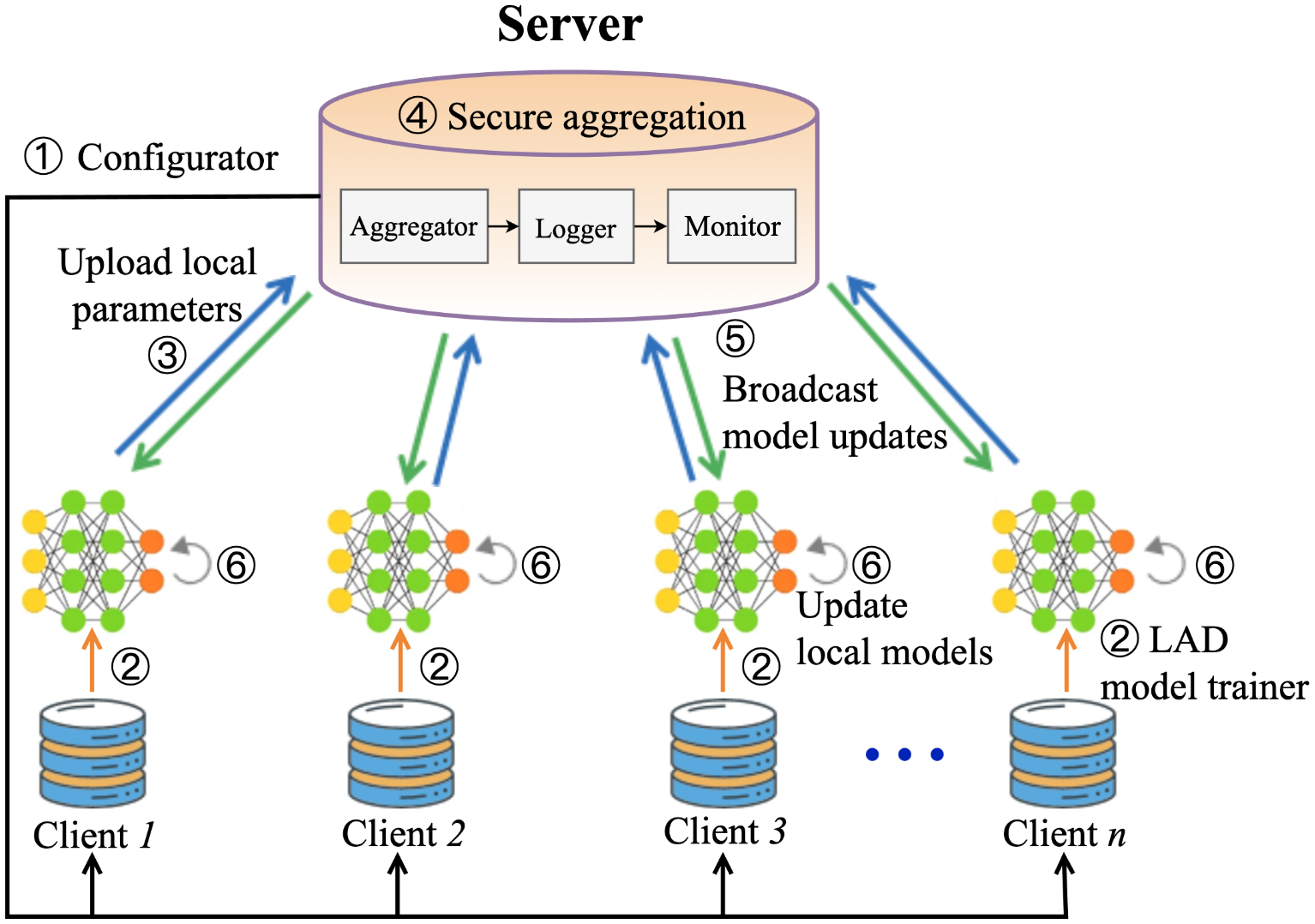}
    \caption{Federated learning workflow in FedLAD.}
    \label{fig:workflow}
\end{figure}

FedLAD adopts a synchronous FL workflow that proceeds in communication rounds, with each round comprising a complete cycle of local training, aggregation, adaptation, and synchronization. This procedural flow, depicted in Fig.\ref{fig:workflow}, illustrates how the Managed System coordinates with the Environment across rounds.

The process begins with \ding{172} where the configurator parses experiment settings from a YAML configuration file, including dataset, model, number of clients, and aggregation strategy.
These configurations are dispatched to both server and client components via the FedLAD API.
In \ding{173}, each simulated client loads its assigned data partition and initializes its corresponding LAD model.
Clients then perform local training for a configurable number of epochs using standard optimizers, independently updating their models based on local log data.
After training, in \ding{174}, clients upload their updated model parameters to the server.
These updates capture each client’s individualized view of system behavior, which is then used to inform global model aggregation and further adaptation logic.

Upon receiving updates, during \ding{175} the server performs global aggregation using the selected strategy (e.g., FedAvg, FedProx, Scaffold).  
During this phase, the Logger captures key performance metrics for each round, such as global validation loss and F1-score.  
In \ding{176}, these metrics are passed to the managing system, which monitors training dynamics and determines whether adaptation should be triggered.  
If predefined criteria are met, such as stagnation in performance over $T$ rounds or a significant drop in accuracy, the Executor may switch aggregation strategies or invoke early stopping to improve training efficiency.  
Finally, in \ding{177}, the aggregated global model is broadcast back to all clients, who update their local models accordingly.  
This completes a full training round, after which the workflow advances to the next iteration.

To summarize, the complete training cycle empowers FedLAD to emulate realistic federated LAD scenarios with dynamic adaptation, offering a flexible and reproducible platform for experimentation across diverse configurations.

%% file: 4_Implementation.tex
\section{Implementation}
\label{sec:implementation}

\subsection{Deployment Environment}

We tested FedLAD on a server equipped with 4× NVIDIA RTX 3090 GPUs (24GB each), running Ubuntu 20.04 and CUDA 11.4. The framework supports Python 3.9 and 3.10. For reproducibility, we recommend a Linux environment with at least 8GB RAM and an NVIDIA GPU (CUDA 11+) for accelerated training. FedLAD supports both single-machine and multi-machine execution through process-based client simulation, with optional GPU acceleration for local training. All components are container-ready, facilitating efficient deployment and end-to-end experimentation.

\subsection{Implementation Details}

\subsubsection{Configuration Management}
FedLAD relies on YAML-based configuration files to specify experiment parameters, such as client count, dataset, model, training epochs, learning rate, aggregation method, and adaptation policy.  
These files are parsed at startup by the configurator, which distributes the settings across system components.  
FedLAD also includes an optional auto-configurator that inspects dataset characteristics to suggest initial hyperparameters, reducing manual tuning.

\subsubsection{Dataset and Model Handling}
FedLAD provides built-in support for benchmark log datasets, including BGL, HDFS, and Thunderbird.  
Raw log files are preprocessed into sequences of event templates using a sliding window mechanism.  
For datasets with external anomaly annotations (e.g., HDFS), label alignment is handled automatically.  
Each client receives a distinct data partition based on the configured IID or non-IID strategy.
LAD models are encapsulated as plug-and-play components following a unified interface.  
This interface defines standard functions for forward propagation, loss computation, and parameter updates, enabling consistent training and aggregation across models.

\subsubsection{Federated Training Engine}
The training engine coordinates server–client communication and implements the full FL protocol.  
Clients train models locally and return state dictionaries, which are aggregated on the server side.  
The engine supports per-client optimizer configurations and simulates heterogeneous local resource conditions.  
All training operations are encapsulated for reusability and extensibility, allowing users to experiment with custom aggregation schemes.  
It leverages PyTorch’s modular training loop to manage local model execution and gradient handling.  
Server-side aggregation is implemented as a strategy-agnostic class, enabling dynamic hot-swapping of optimization methods during runtime.
 
\subsubsection{Aggregation Strategy Implementation}

FedLAD encapsulates aggregation strategies as pluggable components that adhere to a unified interface.  
Each strategy is implemented as an independent Python class defining core aggregation methods.  
During training, the server dynamically loads the selected strategy based on configuration files or adaptation signals from the Executor.  
Strategy selection can be performed statically before training begins or dynamically at runtime, allowing FedLAD to adapt to evolving training conditions.
To support extensibility, new strategies can be integrated by subclassing the base aggregator interface and registering them through the configuration system.
This decoupled design enables researchers to prototype custom aggregation methods within a consistent framework, while FedLAD internally tracks runtime metadata, such as client weights and gradient norms, to support advanced strategies requiring this information.

\subsubsection{Executor and Adaptation Logic}

FedLAD’s Executor enables runtime adaptivity through a monitor–trigger–actuate control loop.  
After each aggregation round, the Logger records validation metrics (e.g., loss, F1-score), which are passed to the Executor for evaluation.  
The Executor checks whether predefined conditions are met, such as performance stagnation over $T$ rounds or F1-score degradation beyond a threshold $\delta$, and triggers appropriate adaptation actions.  
These actions include switching aggregation strategies (e.g., from FedAvg to Scaffold) or invoking early stopping.  
Internally, the Executor operates as a finite state machine that tracks training progress and transitions, ensuring traceable and explainable behavior.

\subsubsection{Logging and Reproducibility}

FedLAD ensures reproducibility by enforcing fixed random seeds across all components and logging comprehensive experiment metadata.  
Recorded logs include per-round metrics, model checkpoints, adaptation events, and client-specific training statistics.  
Results are saved in structured formats (CSV, JSON, PNG) and exported automatically for visualization and analysis.

Together, these components make FedLAD a modular, extensible, and reproducible testbed for FL-based LAD, with extension instructions for models, datasets, and strategies provided in the artifact guide.

%% file: 5_Example_Experiment.tex
\section{Experiments}
\label{sec:experiment}

\subsection{Experiment Workflow}

\begin{figure}[t]
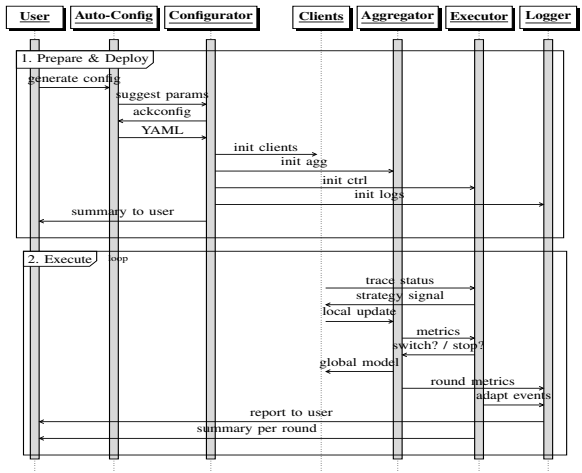

\centering 
\resizebox{0.9\linewidth}{0.7\linewidth}{
\begin{sequencediagram}
    \tikzstyle{every node}=[font=\large]

    \newthread{user}{\textbf{User}} 
    \newthread{auto}{\textbf{Auto-Config}}
    \newthread{config}{\textbf{Configurator}}
    \newinst[1]{client}{\textbf{Clients}} 
    \newthread{agg}{\textbf{Aggregator}}
    \newthread{ctrl}{\textbf{Executor}}
    \newthread{log}{\textbf{Logger}}

    \begin{sdblock}{\large 1. Prepare \& Deploy}{}
        \mess{user}{generate config}{auto}
        \mess{auto}{suggest params}{config}
        \mess{config}{ackconfig}{auto}
        \mess{auto}{YAML}{config}
        \mess{config}{init clients}{client}
        \mess{config}{init agg}{agg}
        \mess{config}{init ctrl}{ctrl}
        \mess{config}{init logs}{log}
        \mess{config}{summary to user}{user}
    \end{sdblock}

    \begin{sdblock}{\large 2. Execute}{loop}
        \mess{client}{trace status}{ctrl}
        \mess{ctrl}{strategy signal}{client}
        \mess{client}{local update}{agg}
        \mess{agg}{metrics}{ctrl}
        \mess{ctrl}{switch? / stop?}{agg}
        \mess{agg}{global model}{client}
        \mess{agg}{round metrics}{log}
        \mess{ctrl}{adapt events}{log}
        \mess{log}{report to user}{user}
        \mess{ctrl}{summary per round}{user}
    \end{sdblock}

\end{sequencediagram}}
\caption{Two-stage lifecycle in FedLAD.}
\label{fig:two-stage}
\end{figure}

To demonstrate how developers can use FedLAD to evaluate their own LAD models or federated training strategies, this section presents a full experiment workflow composed of two stages: prepare and deploy, and execute, as illustrated in Fig.~\ref{fig:two-stage}. The overall lifecycle enables pluggable customization and adaptive evaluation in FL.

\textbf{Preparation and Deployment.}  
The user starts by registering their LAD model or aggregation strategy into the corresponding registry. FedLAD provides standardized interfaces for custom extensions, allowing researchers to evaluate new algorithmic designs without modifying the underlying infrastructure. After implementation, users may define manually configuration parameters or invoke the built-in assistant script, which automatically recommends dataset-specific settings. These settings are passed to the Configurator and finalized into a YAML file. The system then initializes the full pipeline, including client simulators, the Aggregator, the Executor for adaptive logic, and the logger for tracking performance. A summary is returned for user validation. The deployment phase completes with all clients assigned data partitions (IID or non-IID) and a training environment for execution.

\textbf{Execution.}  
In the execution stage, each round proceeds by simulating local training at each client. Clients report runtime status to the Executor, which evaluates current progress and may issue adaptive instructions such as switching strategies or triggering early stopping. Local model updates are sent to the Aggregator, which performs federated aggregation (e.g., FedAvg or a user-defined strategy) and broadcasts the global model back to clients. Round-wise metrics, including validation loss and F1-score, are logged in real-time. Throughout the process, the Executor continuously evaluates whether adaptation criteria are met, such as plateaued performance, and triggers corresponding events. Logs of all metrics, events, and decisions are reported to the user. After training concludes, FedLAD exports all relevant artifacts, including model checkpoints, metric logs, and adaptation history to support analysis.

This end-to-end workflow demonstrates how researchers can use FedLAD to easily test, compare, and analyze their customized algorithms in realistic federated LAD scenarios, benefiting from the testbed’s modular design, automation capabilities, and adaptive control mechanisms.

\subsection{Experiment Results}
To illustrate the evaluations of FedLAD, this subsection reports example results from running an LAD model on a benchmark dataset under a federated setup with 50 simulated clients. After specifying the model and training configurations, researchers can use FedLAD's built-in logging and visualization modules to obtain detailed runtime insights.

\begin{figure}[htbp]
    \centering
    \begin{subfigure}[b]{0.45\textwidth}
        \centering
        \includegraphics[width=\textwidth]{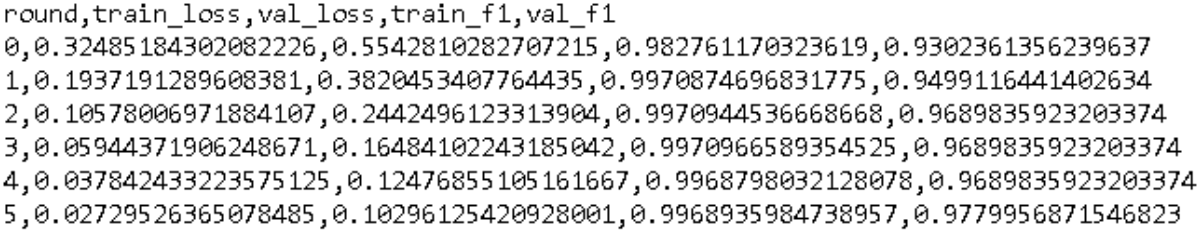}
        \caption{Per-Round Metrics Tracked by the Training Monitor.}
        \label{fig:train_result}
    \end{subfigure}
    
    \begin{subfigure}[b]{0.24\textwidth}
        \centering
        \includegraphics[width=\textwidth]{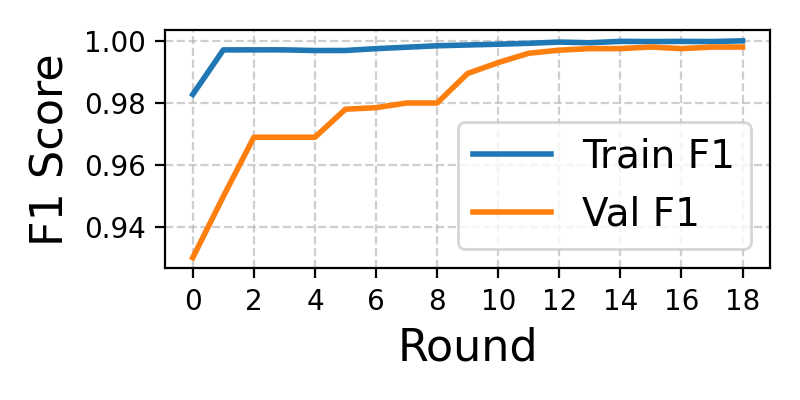}
        \caption{F1-score over rounds}
        \label{fig:f1_curve}
    \end{subfigure}
    \hfill
    \begin{subfigure}[b]{0.24\textwidth}
        \centering
        \includegraphics[width=\textwidth]{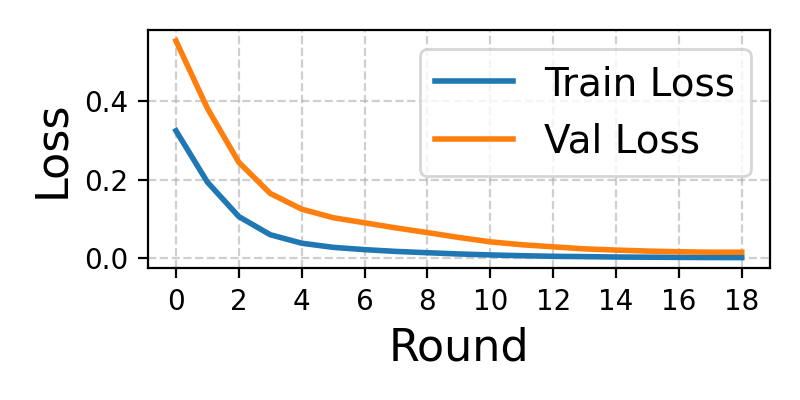}
        \caption{Loss over rounds}
        \label{fig:loss_curve}
    \end{subfigure}
    \caption{Detection performance in FedLAD. Parameter settings: \textit{dataset} = Thunderbird, \textit{distribution} = non-IID, \textit{model} = LogAnomaly.}
    \label{fig:detection_performance}
\end{figure}

\textbf{Detection Performance.}  
Fig.~\ref{fig:detection_performance} shows the detection performance curve of LogAnomaly under a non-IID data distribution, illustrating how performance varies across training rounds in the absence of adaptation. The training monitor captures key performance indicators such as F1-score and loss at each round. These results are automatically logged in CSV format for further analysis. Developers can easily visualize convergence behavior across different configurations to compare training stability and model generalization. In this example, the model rapidly achieves an F1-score above 0.98 within two rounds and converges toward 0.997, demonstrating strong detection performance with minimal training cost.

\begin{figure}
    \centering
    \includegraphics[width=0.55\linewidth]{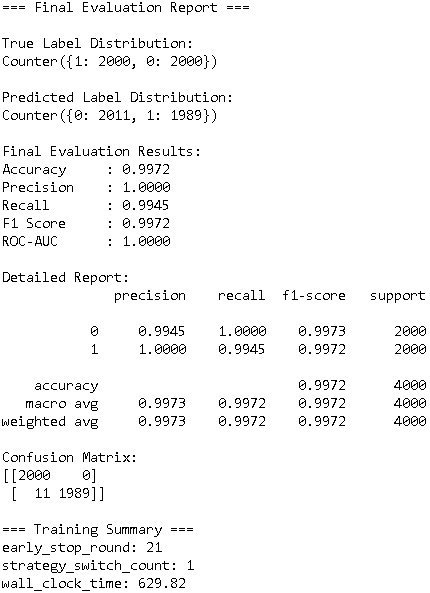}
    \caption{Example of the final evaluation report with adaptive control enabled. Parameter settings: \textit{dataset} = Thunderbird, \textit{distribution} = IID, \textit{model} = NeuralLog.}
    \label{fig:txt}
\end{figure}

To further demonstrate FedLAD’s versatility, we compare centralized and FL training using NeuralLog on the Thunderbird dataset. Existing FL-based LAD implementations are often tightly integrated with specific models or datasets, lacking standardized support for evaluating alternative architectures or training setups such as centralized LAD. The centralized setup processes all logs on a single client, while the FL distributes data across ten clients. Both achieve comparable accuracy (0.9977 vs. 0.9992), with the FL variant slightly outperforming in recall and F1-score. Training time remains similar (80.2s vs. 90.1s), and the communication overhead is moderate, which is about 27.5MB per client across 11 rounds. These results show that FedLAD supports FL workflows with minimal overhead, validating its extensibility and practical efficiency. 

FedLAD’s adaptive Executor empowers researchers to evaluate runtime decision-making under dynamic conditions. Fig.~\ref{fig:txt} illustrates the execution trace for NeuralLog under an IID setting, where the adaptive mechanism enabled both early stopping and strategy switching. Specifically, the Executor halted training at round 21 upon detecting a validation, reducing unnecessary computation. Additionally, it triggered a strategy switch in response to performance drift, demonstrating how FedLAD supports real-time control adaptations without entangling model logic. Besides, total training time was approximately 630 seconds, including model updates, aggregation, and adaptation overheads. The high F1-score of 0.9972 and rapid convergence underscore FedLAD’s utility in supporting fast FedLAD evaluation. 

To evaluate the benefits of these adaptive features, we compared them with a fixed-strategy baseline using the same model and dataset but without early stopping or strategy switching. The adaptive setup achieved a higher F1-score (0.9972 vs. 0.9946), over 25\% less training time, fewer communication rounds, and better convergence. These results highlight the practical advantage of FedLAD’s self-adaptation over fixed-policy FL training, suggesting that adaptive mechanisms improve efficiency and generalization by mitigating overfitting.


%% file: 6_Threat_to_Validity.tex
\section{Threat to Validity}
While FedLAD provides a flexible and modular testbed for benchmarking federated LAD models, it also has certain limitations. First, FedLAD currently simulates clients on a single machine, which may not fully reflect network communication overheads in real-world FL deployments. Secondly, although we support three public datasets, expanding to industrial-scale logs may require custom parsers and encoders. Finally, the adaptation controller currently focuses on early stopping and aggregation switching, where future work can explore more advanced control strategies such as reinforcement learning.

%% file: 7_Conclusion.tex
\section{Conclusion and Future Works}
\label{sec:conclusion}

This paper presented FedLAD, a modular and self-adaptive testbed for benchmarking LAD in federated environments.  
FedLAD supports realistic and reproducible experimentation by enabling pluggable LAD models, benchmark log datasets, configurable IID/non-IID data partitioning, and multiple aggregation strategies.  
Its modular architecture separates configuration, coordination, and execution concerns, supporting extensibility, scalability, and reproducibility.
FedLAD further integrates self-managing features, including self-monitoring for logging per-round metrics, self-configuration for automatic hyperparameter tuning, and self-adaptation for enabling runtime decisions such as strategy switching and early stopping.
By lowering the barrier to prototyping, evaluation, and extension, FedLAD provides a unified platform for advancing privacy-preserving, scalable, and adaptive LAD research under federated constraints.
Future work will explore deployment across real-world distributed agents, integration of advanced privacy defenses, and support for upstream log preprocessing tasks, including automated log parsing.